\documentclass[sigconf, natbib=true]{acmart}
\renewcommand\footnotetextcopyrightpermission[1]{}

\AtBeginDocument{%
  }
\acmConference[CIKM26]{ACM CIKM Conference on Knowledge and Information Management}{November 7--11, 2026}{Rome}

\usepackage{array}
\usepackage{url}
\usepackage[most]{tcolorbox}
\usepackage{subcaption}
\usepackage{pifont}
\newcommand{\cmark}{\textcolor{green}{\ding{51}}}%
\newcommand{\xmark}{\textcolor{red}{\ding{55}}}%

\begin{document}

\title{When Is Complex Chunking Worth It? A Multi-Objective Evaluation of Chunking Methods at Scale}

\author{Laura Caspari}
\authornote{Corresponding author}
\orcid{0009-0002-6670-3211}
\affiliation{%
  \department{Chair of Data Science}
  \institution{University of Passau}
  \streetaddress{Innstraße 33}
  \city{Passau}
  \country{Germany}
  \postcode{94032}
}
\email{laura.caspari@uni-passau.de}

\author{Kanishka Ghosh Dastidar}
\email{kanishka.ghosh-dastidar@it-u.at}
\orcid{0000-0003-4171-0597}
\affiliation{%
  \institution{Interdisciplinary Transformation University}
  \streetaddress{Innstraße 33}
  \city{Linz}
  \country{Austria}
  \postcode{94032}
}

\author{Michael Dinzinger}
\email{michael.dinzinger@uni-passau.de}
\orcid{0009-0003-1747-5643}
\affiliation{%
  \institution{University of Passau}
  \streetaddress{Innstraße 33}
  \city{Passau}
  \country{Germany}
  \postcode{94032}
}

\author{Jelena Mitrovi\'{c}}
\email{jelena.mitrovic@uni-passau.de}
\orcid{0000-0003-3220-8749}
\affiliation{%
  \institution{University of Passau}
  \streetaddress{Innstraße 33}
  \city{Passau}
  \country{Germany}
  \postcode{94032}
}

\author{Michael Granitzer}
\email{michael.granitzer@uni-passau.de}
\orcid{0000-0003-3566-5507}
\affiliation{%
  \institution{University of Passau}
  \streetaddress{Innstraße 33}
  \city{Passau}
  \country{Germany}
  \postcode{94032}
}
\affiliation{%
  \institution{Interdisciplinary Transformation University}
  \streetaddress{Innstraße 33}
  \city{Linz}
  \country{Austria}
  \postcode{94032}
}

\renewcommand{\shortauthors}{Laura Caspari, Kanishka Ghosh Dastidar, Michael Dinzinger, Jelena Mitrovi\'{c}, Michael Granitzer}

\begin{abstract}
Dense retrieval is commonly evaluated on benchmarks that represent each document with a single embedding, even though real-world retrieval systems often index long documents that require chunking.
In these settings, the chosen chunking method not only affects retrieval quality, but also indexing throughput, query latency, and memory usage.
Prior comparisons of chunking strategies have mainly focused on retrieval performance, leaving operational trade-offs underexplored.
To address these issues, we evaluate eight representative chunking strategies across two scalable corpora, three embedding models, and multiple corpus sizes, measuring both retrieval effectiveness and system-level costs.
Our results show that computationally expensive methods rarely provide consistent gains over simpler chunking.
Instead, the best performing strategy depends on the embedding model, dataset, corpus size, and target retrieval metric.
Methods with similar performance can also differ substantially in operational cost, showing that chunking should be seen as a multi-objective design decision.
Our code and datasets are available on GitHub\footnote{\url{https://github.com/casparil/chunking-eval}} and Hugging Face\footnote{\url{https://huggingface.co/datasets/PaDaS-Lab/kilt-nq}}\footnote{\url{https://huggingface.co/datasets/PaDaS-Lab/CoRE}}.
\end{abstract}




\maketitle

\begingroup
  \renewcommand{\thefootnote}{}
  \footnotetext{%
    The definitive Version of Record was published in \textit{CIKM '26: Proceedings of the 35th ACM International Conference on Information and Knowledge Management}, 2026. \url{https://doi.org/10.1145/3799682.3839948}.
  }
  \addtocounter{footnote}{-1}
\endgroup

\section{Introduction}

Retrieval-Augmented Generation (RAG) has become a key component of modern retrieval pipelines, often using dense embedders to retrieve additional query relevant information~\cite{Yu2024-arxiv}.
The performance of these systems is commonly evaluated on benchmarks like the Massive Multilingual Text Embedding Benchmark (MMTEB)~\cite{Enevoldsen2025}, which provide standardized retrieval tasks across languages, domains, and datasets.
However, documents are generally represented by a single embedding with long inputs being truncated to fit into the model's context window.
While this simplifies evaluation, deployed collections often contain long web pages, manuals or PDFs that must be split into smaller units.
Chunking therefore becomes a core design choice in dense retrieval pipelines that determines how much context is preserved in each indexed unit, and how many embeddings are created per document.
Recent work has proposed a range of strategies \cite{Kamradt2024, Anthropic2024, Guenther2024-arxiv, Gao2023-arxiv} from simple token-based splitting to methods relying on large language models (LLMs) that differ substantially in computational cost.

Despite this, prior comparisons of chunking methods have mostly focused on retrieval performance alone \cite{Merola2025, Qu2025, Ruke2025}, leaving answers to two practical questions lacking.
First, do more complex chunking methods provide retrieval gains that justify their additional cost?
Second, does the answer change as the corpus grows?
The second question is of particular importance given that dense retrieval performance deteriorates as collections grow~\cite{Reimers2021, Caspari2025}, potentially favoring approaches that generate fewer chunks.

\begin{table*}[t!]
    \centering
    \caption{List of compared chunking methods with a short description.}
    \vspace{-1em}
    \label{tab:chunking}
    \begin{tabular}{p{0.15\linewidth}p{0.7\linewidth}>{\centering\arraybackslash}p{0.1\linewidth}}
        \toprule
        Method & Description & Additional Model Calls \\ \midrule
        Token & Fixed-size token windows with overlap. & \xmark \\
        Sentence & Token windows adjusted to end at sentence boundaries. & \xmark \\
        Late \cite{Guenther2024-arxiv} & Embeds the full document and chunks unpooled word embeddings before mean-pooling. & \xmark \\
        Enriched (Title) & Prepends document title metadata to each chunk. & \xmark \\
        Enriched (Summary) & Prepends a document summary to each chunk. & \cmark \\
        Contextual \cite{Anthropic2024} & Prepends LLM-generated, chunk-specific context that situates the chunk in the full document. & \cmark \\
        Summary & Represents each document with a generated summary. & \cmark \\
        Semantic & Splits documents into sentence groups using sentence-embedding similarity. & \cmark \\
        \bottomrule
    \end{tabular}
\end{table*}

We address these questions through a multi-objective evaluation of eight representative chunking methods on two scalable corpora, using three open-source embedding models.
In addition to retrieval performance, we measure document throughput, query throughput and peak memory usage.
This setup allows us to study not only which methods retrieve better, but also whether their gains are robust across models, datasets, corpus sizes, and operational constraints.

Our results show that computationally expensive chunking methods rarely outperform simpler alternatives, with the best-performing strategy differing across evaluation settings.
Moreover, methods with similar retrieval scores can differ substantially in indexing throughput, query throughput, and memory requirements.
These findings suggest that chunking should be treated as a multi-objective design decision: the appropriate method depends on the target retrieval metric as well as the system constraints of the intended application.

\section{Related work}

\paragraph{Chunking Strategies}
Chunking refers to the process of breaking down input text into smaller, manageable pieces~\cite{Smith2024}.
In its simplest form, chunking uses fixed-sized token windows to split text, optionally with overlap.
As this can cause chunk boundaries to occur mid-sentence, sentence-based chunking attempts to end each chunk at natural boundaries, while structure-aware methods exploit document hierarchies to find meaningful chunks \cite{Nguyen2025, Shin2025}.
Late chunking \cite{Guenther2024-arxiv} leverages the long context of modern embedders to process the entire document, then split on the generated word embeddings before pooling.
More advanced strategies employ language models to define chunk boundaries \cite{Kamradt2024, Duarte2024, Jain2025} or use them to enrich chunks with additional information \cite{Anthropic2024}.
Graph-based RAG methods provide another way to preserve relationships across document fragments and have been shown to improve retrieval in some settings \cite{Rani2024, Edge2025}.

\paragraph{Comparison of Chunking Methods}
As chunking can strongly influence retrieval behavior, the comparison of chunking strategies has received increasing attention \cite{Anthropic2024, Guenther2024-arxiv, Duarte2024, Jain2025, Ruke2025, Qu2025, Merola2025}.
However, most studies emphasize retrieval effectiveness and report very limited evidence on operational costs.
To the best of our knowledge, the work closest to this paper is \cite{Qu2025}, which explicitly studies whether the performance of semantic chunking justifies its computational cost.
Our work extends this perspective by comparing a broader set of chunking strategies across models and scalable corpora.

\paragraph{Corpus Scale}
Our scaling analysis is motivated by prior evidence that dense retrievers can degrade as corpus size increases \cite{Reimers2021, Caspari2025}.
This issue might interact with chunking due to the different number of indexed vectors generated by different methods.
While methods that create many chunks can make short query-relevant passages easier to retrieve \cite{Bhat2025, Chen2024}, the overall larger index size might also make retrieval more difficult as the corpus grows.

\section{Methodology}

We evaluate chunking strategies under both retrieval-effectiveness and system-efficiency criteria. For each dataset, corpus size, embedding model, and chunking method, we chunk the corpus, embed all resulting units, store them in a FAISS index \cite{Douze2024}, and retrieve the top-ranked chunks for each query. Chunk-level retrieval scores are mapped back to documents for evaluation. We report NDCG@10 to measure single-stage retrieval quality and Recall@100 to measure first-stage retrieval performance when a downstream reranker or generator may consume a larger candidate set.

\subsection{Chunking Strategies}
\label{chunking}
Table \ref{tab:chunking} lists the compared chunking methods together with a short description.
The additional model calls column specifies whether the method requires additional calls to the embedding model or an external LLM.
For token, sentence, late, enriched, and contextual chunking, we use 512-token chunks with an overlap of 25 tokens before adding optional metadata or generated context.
Semantic chunking embeds sentences with the same embedding model used for retrieval and starts a new chunk when similarity falls below the 95th percentile threshold.
For the summary and contextual chunking, we use a local Qwen3-Next-80B-A3B-Instruct model \cite{Qwen32025, Yang2025} quantized to 8 bits.
Generated outputs are produced before indexing and are reused across embedding models where applicable.

\subsection{Embedding Models}
We evaluate three open-source embedding models below 1B parameters: Qwen-0.6B \cite{Qwen32025}, embeddinggemma-300M \cite{Vera2025}, and Snowflake-L V2 \cite{Yu2024}. These models differ in parameter count, embedding dimensionality, and MMTEB retrieval performance.

\begin{figure*}[t!]
    \centering
    \includegraphics[width=0.9\linewidth,trim={0 0 0 2.7cm},clip]{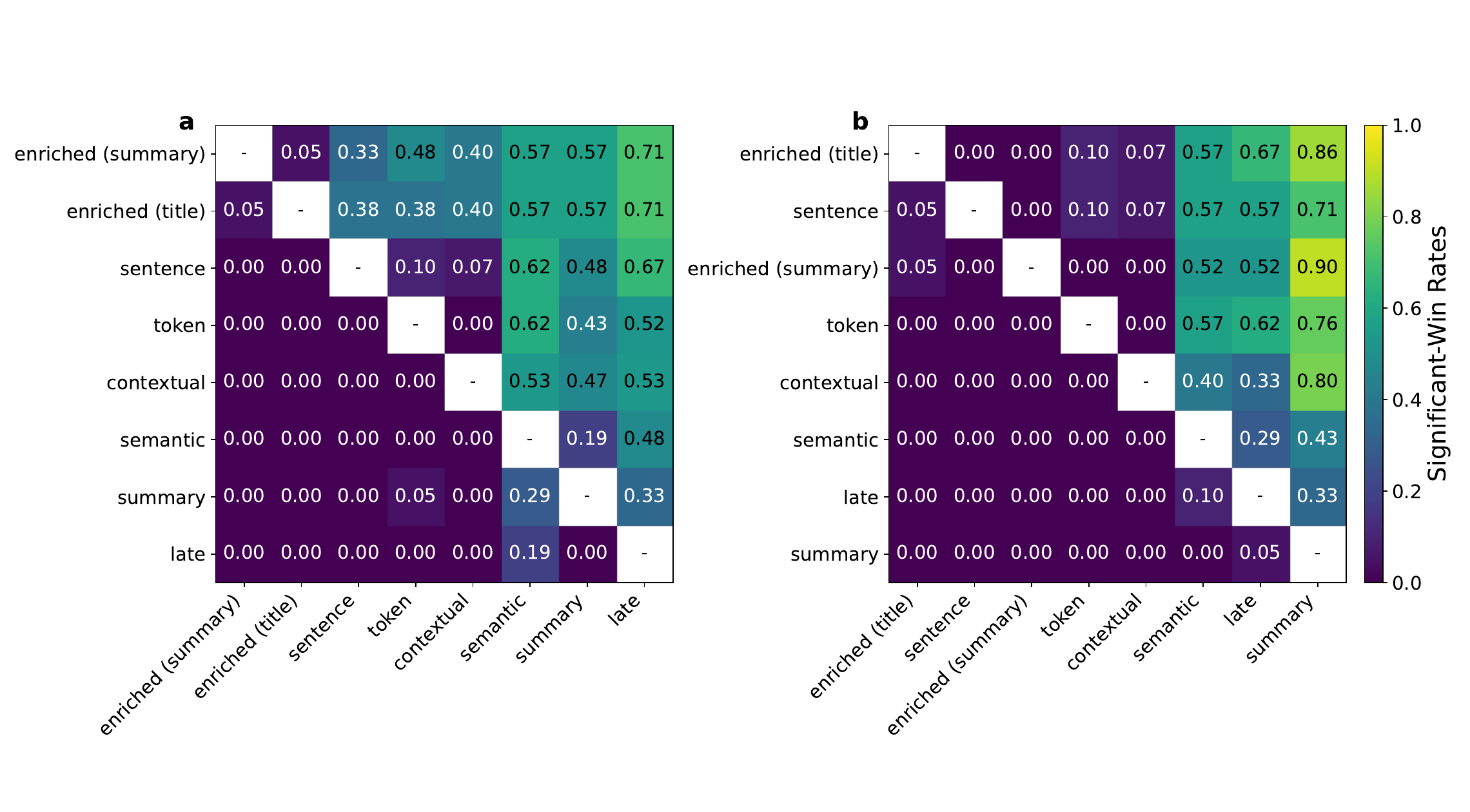}
    \vspace{-3em}
    \caption{\emph{Significant-win rates across all corpora and models for NDCG@10 (a) and Recall@100 (b)}. Cell \texttt{(i, j)} is the proportion of settings in which method \texttt{i} significantly outperforms method \texttt{j}. For example, the score of \texttt{0.71} in the upper right corner of (a) states that enriched (summary) chunking outperforms late chunking 71\% of the time.}
    \label{fig:dominance}
    \vspace{-1em}
\end{figure*}

\subsection{Datasets}
We use two scalable corpora: The CoRE document corpus \cite{Caspari2025} and KILT \cite{Petroni2021} with NQ relevance judgements and queries \cite{Kwiatkowski2019}.
CoRE contains four corpora of distinct sizes (10K, 100K, 1M, 10M) derived from MS MARCO v2\footnote{\url{https://github.com/zhouyonglong/MSMARCOV2}}.
Following the same construction strategy used for CoRE, we create KILT subcorpora of 10K, 100K, 1M, and 6M documents\footnote{\url{https://huggingface.co/datasets/PaDaS-Lab/kilt-nq}}.
Since KILT only contains around 6M documents in total, this is the largest available corpus.
For evaluation, we chunk the entire KILT scale but restrict chunking on CoRE to 1M documents, as this already produces around 5M indexed embeddings depending on the method.
Furthermore, we only scale the contextual chunking approach to 100K on CoRE and 1M on KILT due to its high computational cost.

\subsection{Runtime Analysis}
In addition to retrieval effectiveness, we measure document throughput during indexing, query throughput during retrieval, and peak memory usage.
These metrics capture different operational bottlenecks with document throughput reflecting corpus ingestion cost, query throughput serving latency, and memory usage deployment constraints.

\subsection{Pairwise Significance Tests}
\label{subsec:pairwise}
For each model, dataset, and corpus size, we compare chunking methods using query-level retrieval scores.
In each of these settings, i.e. for a fixed model, dataset, and corpus size, we apply Fisher's randomization test \cite{Fisher1936} with 10,000 permutations and Bonferroni correction over the 28 pairwise comparisons of chunking methods.
The resulting pairwise tests are summarized as significant-win rates across settings.

\section{Results}

\begin{table*}[t!]
    \centering
    \caption{Comparison across models, datasets and corpus sizes for Recall@100. The top-performing method is marked in bold, the second-best is underlined.}
    \label{tab:top}
    \vspace{-1em}
    \begin{tabular}{lcccc||cccc||cccc}
            \toprule
            & \multicolumn{4}{c}{Gemma} & \multicolumn{4}{c}{Qwen} & \multicolumn{4}{c}{Snowflake} \\
            & \multicolumn{2}{c}{CoRE} & \multicolumn{2}{c}{KILT} & \multicolumn{2}{c}{CoRE} & \multicolumn{2}{c}{KILT} & \multicolumn{2}{c}{CoRE} & \multicolumn{2}{c}{KILT} \\
            \midrule
            Method & 10K & 1M & 10K & 1M & 10K & 1M & 10K & 1M & 10K & 1M & 10K & 1M \\
            \midrule
            Token & \textbf{82.73} & \underline{57.09} & 98.82 & 94.68 & 76.73 & \textbf{56.73} & 98.31 & 94.03 & 77.64 & 54.18 & \textbf{98.83} & 92.87 \\
            Sentence & \underline{82.55} & 56.00 & 98.83 & \underline{94.91} & 76.18 & \underline{55.82} & \textbf{98.36} & \underline{94.04} & \textbf{79.09} & \textbf{55.27} & 97.90 & 93.23 \\
            Enriched (Title) & \underline{82.55} & \textbf{57.27} & \underline{98.86} & \textbf{94.95} & \textbf{78.00} & 55.27 & \textbf{98.36} & \textbf{94.28} & 78.18 & 54.73 & 97.96 & 93.13 \\
            Late & 80.18 & 50.54 & 98.71 & 94.00 & 75.46 & 48.00 & 97.79 & 88.79 & 76.91 & 48.36 & 97.84 & 89.11 \\
            \midrule
            Semantic & 80.00 & 48.00 & 98.56 & 93.51 & 75.27 & 52.36 & 98.16 & 93.10 & 77.09 & 46.91 & 97.72 & 91.39 \\
            Summary & 79.82 & 51.82 & 98.46 & 92.86 & 71.64 & 46.36 & 96.80 & 87.61 & 72.36 & 46.73 & 96.84 & 88.67 \\
            Enriched (Summary) & 82.36 & 55.64 & \textbf{98.90} & 94.67 & 77.45 & 54.91 & 98.33 & 93.81 & \underline{78.91} & \underline{55.09} & \underline{98.07} & \textbf{93.26} \\
            Contextual & 82.18 & - & 98.85 & 94.30 & \underline{77.82} & - & \underline{98.35} & 93.93 & \underline{78.91} & - & 97.84 & \underline{93.25} \\
        \bottomrule
    \end{tabular}
\end{table*}

\paragraph{Simple Methods Remain Competitive.}
Figure \ref{fig:dominance} summarizes pairwise significance tests as significant-win rates across models, datasets, and corpus sizes.
Each cell shows how often the method in the row significantly outperforms the method in the column.
The results do not support the hypothesis that more complex chunking methods consistently improve retrieval effectiveness.
In particular, semantic chunking, summary-only indexing, contextual chunking, and late chunking rarely achieve significant wins over simpler baselines.
The strongest exception is enriched (summary) chunking for NDCG@10 (a), which often improves over several alternatives, suggesting that document-level context can help with single-stage retrieval.
However, this pattern is not universal: enriched (summary) chunking rarely wins over enriched (title) chunking, and its advantage is weaker under Recall@100 (b).
Overall, the significant-win rates indicate robust underperformance for some expensive methods, but do not identify a single best method across settings.

\paragraph{Win Rates Change with Retrieval Metric.}
The comparison between NDCG@10 (a) and Recall@100 (b) in Figure \ref{fig:dominance} shows that choosing an appropriate chunking method can depend on the intended retrieval use case.
With NDCG@10, enriched methods are often stronger as they seem to be more effective at improving the ordering of the highest-ranked documents.
When using Recall@100, token and sentence chunking become much more competitive, making them a viable choice for first-stage retrieval.
A method that is suboptimal for single-stage ranking can therefore still be appropriate for first-stage retrieval.

\paragraph{Corpus Size, Dataset and Model Matter}
As Table \ref{tab:top} illustrates, the best method varies across models, datasets, and corpus sizes with no chunking strategy consistently outperforming the rest.
The observations are in line with win rates shown in Figure \ref{fig:dominance} (b), with enriched (title) chunking performing well in most settings, and token or sentence baselines being close competitors.
Late chunking and summary-only indexing are often weaker, especially at larger corpus sizes.
While the table only shows results for certain corpus sizes, the described trends transfer to unshown settings.
For reference, the full result tables are available in our repository\footnote{\url{https://github.com/casparil/chunking-eval/blob/main/results.md}}.

While not evident from the table, we observe an interesting pattern between win rates and corpus size.
The number of significant differences between chunking methods generally increases as the corpus grows, suggesting that chunking becomes more consequential in larger indexes.
However, the trend is not perfectly monotonic, as the largest 6M KILT corpus has fewer significant wins than the 1M corpus.

\begin{figure}[b!]
    \centering
    \includegraphics[width=\linewidth]{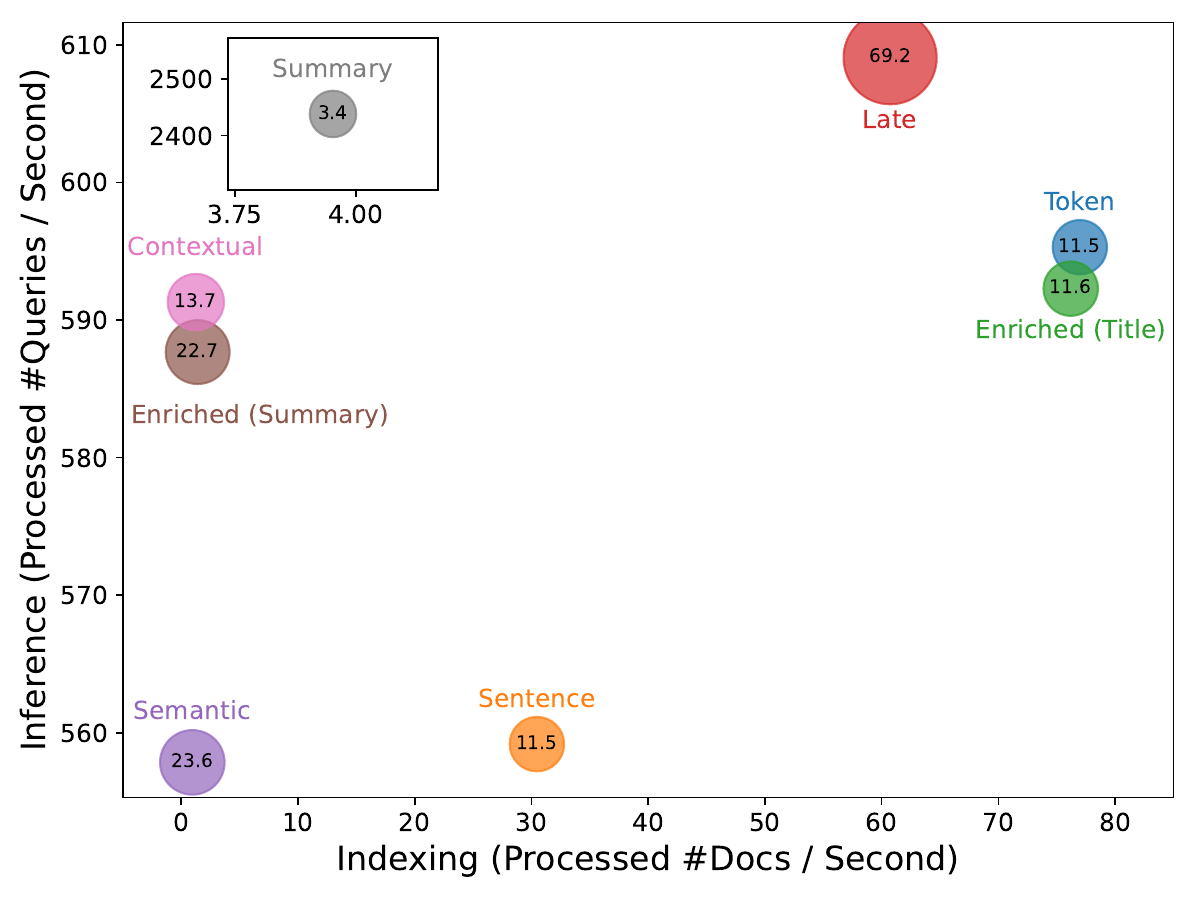}
    \vspace{-2em}
    \caption{\emph{Average query and document throughputs per chunking method for Gemma on KILT 10k.} The circles represent the method's RAM usage (in GB) at index construction time. Summary-only chunking is shown in its own internal plot, as its much higher query throughput would otherwise obscure differences between the remaining methods.}
    \label{fig:pareto}
    \vspace{-1em}
\end{figure}

\paragraph{Runtime Analysis Allows Optimization Beyond Pure Performance.}
Figure \ref{fig:pareto} reports a representative runtime comparison for Gemma on KILT 10K, illustrating that methods with similar retrieval effectiveness can differ substantially in document throughput, query throughput, and memory use.
Token chunking has high indexing throughput and relatively low memory usage, while sentence chunking is slower despite being conceptually simple.
Semantic, contextual, and summary-based methods are much slower to index as they require additional embedding or LLM generation steps.

The observed query throughput is closely tied to index size.
Summary-only indexing creates one representation per document and is therefore fast at query time, at the cost of lower retrieval effectiveness and expensive document processing.
Late chunking has high memory usage during indexing as unpooled token representations are temporatily kept in memory before chunk embeddings are finalized.
These results show that considering retrieval performance alone is insufficient when choosing a chunking method, since a method that is marginally better in NDCG@10 may be unattractive if it substantially increases indexing time, memory usage, or inference cost.

A key factor that heavily influences the efficiency of LLM-based chunking methods is the average number of tokens the LLM can produce per second.
Table \ref{tab:throughput} illustrates how the number of processed documents changes depending on generation throughput, remaining far behind simpler methods even at large numbers.
This makes it difficult to justify the usage of these methods for large corpora unless their retrieval gains are substantial or the corpus is small enough that preprocessing cost is acceptable.
In addition, while we conducted the experiments with a local LLM, using a third-party model can lead to substantial costs.
As an example, using OpenRouter prices as basis for our calculation leads to an estimated cost of 9.6-14.73\$ (provider-dependent) for contextual chunking on the KILT 10K corpus at the time of writing when querying a Qwen3-Next-80B-A3B-Instruct model.
While prefix-caching can reduce these prices, chunking larger corpora with this method would still incur significant cost.

\paragraph{Practical Implications}
For most large-scale retrieval settings, token-based chunking remains a strong default due to is simplicity, fast indexing, memory efficiency and often competitive performance.
Sentence-based chunking presents a reasonable alternative when preserving sentence boundaries is important, but its lower document throughput should be considered.
If document titles are available, enriched (title) chunking is a low-cost improvement that performs competitively across many settings.
While enriched (summary) chunking has shown strong performance, it increases the costs of index construction and should be validated against a cheaper enriched baseline.
Contextual, semantic, late, and summary-only methods should be treated as specialized choices rather than defaults as, based on our experiments, their additional cost is rarely matched by retrieval gains.

In practice, the choice of chunking method should follow the retrieval objective.
For single-stage retrieval, enriched chunking methods should be compared to simpler baselines.
For first-stage retrieval, simple token or sentence chunks may be sufficient.
For large or frequently updated corpora, indexing throughput and memory usage should be considered early, as expensive chunking methods might make re-indexing impractical.

\begin{table}[t!]
    \centering
    \caption{Number of processed documents per second depending on the average number of generated output tokens.}
    \label{tab:throughput}
    \vspace{-1em}
    \begin{tabular}{@{}ccccc@{}}
        \toprule
        & \multicolumn{4}{c}{\# Tokens/s} \\
        Method & 100 & 200 & 500 & 2000 \\ \midrule
        Contextual & 0.26 & 0.53 & 1.31 & 5.26 \\
        Summary & 0.77 & 1.57 & 3.93 & 15.74 \\
        \bottomrule
    \end{tabular}
\end{table}

\section{Conclusion and Limitations}

We evaluated eight chunking strategies across two scalable corpora and three embedding models, considering both retrieval effectiveness and operational costs.
Our results show that computationally expensive methods rarely provide consistent gains over simpler baselines.
Instead, the best method depends on the retrieval metric, embedding model, dataset, corpus size, and deployment constraints.

This study has several limitations.
First, runtime and memory measurements are dependent on our concrete implementation, hardware, batching strategy, and FAISS configuration.
Thus, they should be interpreted as comparative measurements within one controlled setup rather than universal constants.
Second, we evaluate on two scaled retrieval corpora, allowing us to study scaling behavior but being limited in document types, domains, and query styles.
Third, some expensive methods could not be evaluated at the largest corpus sizes, so their large-scale behavior is only partially observed.
Finally, we use fixed chunking hyperparameters rather than tuning each method separately, which improves comparability but may understate the best possible performance of individual methods.

Overall, our findings suggest that chunking should be treated as a multi-objective design choice. Simple methods are strong defaults, while more expensive strategies should be justified by measured gains under the target metric and deployment constraints.

\section{GenAI Usage Disclosure}

The authors used GenAI to edit and improve texts written as part of this paper. All AI-generated text was carefully reviewed prior to submission. In addition, GenAI-assisted code support tools were employed during development. Specific implementation-related questions were also posed to chat-based GenAI models, and the responses were used as a basis for adapting the code accordingly.

\begin{acks}
This work was funded by the Bavarian State Ministry of Economic Affairs, Regional Development, and Energy (StMWi).
\end{acks}

\bibliographystyle{ACM-Reference-Format}
\bibliography{sources}

\end{document}